\documentclass[epj]{webofc}
\usepackage[varg]{txfonts}   
\usepackage{hyperref}

\renewcommand*{\eqref}[1]{Eq.~(\ref{eq:#1})}

\wocname{ARENA-2016}

\usepackage{amsmath}
\usepackage{amssymb}
\usepackage[separate-uncertainty=true,
 range-units = single,
 range-phrase = --
]{siunitx}

\DeclareSIUnit\parsec{pc}
\DeclareSIUnit\lightyear{ly}
\DeclareSIUnit\gauss{G}
\DeclareSIUnit\Sigma{$\sigma$}
\DeclareSIUnit\year{yr}
\DeclareSIUnit\years{yr}
\DeclareSIUnit\erg{erg}

\usepackage[draft]{fixme}
\woctitle{ARENA-2016}
\begin{document}
\title{Search for Cosmic Particles with the Moon and LOFAR}
%
%



\author{
\firstname{T.} \lastname{Winchen}\inst{1}\fnsep\thanks{\email{tobias.winchen@rwth-aachen.de}} 
\and\firstname{A.} \lastname{Bonardi}\inst{2}
\and\firstname{S.} \lastname{Buitink}\inst{1}
\and\firstname{A.} \lastname{Corstanje}\inst{2}
\and\firstname{J. E.} \lastname{Enriquez}\inst{2}
\and\firstname{H.} \lastname{Falcke}\inst{2,5,7}
\and\firstname{J. R.} \lastname{Hörandel}\inst{2}
\and\firstname{P.} \lastname{Mitra}\inst{1} 
\and\firstname{K.} \lastname{Mulrey}\inst{1} 
\and\firstname{A.} \lastname{Nelles}\inst{3} 
\and\firstname{J. P. } \lastname{Rachen}\inst{2}
\and\firstname{L.} \lastname{Rossetto}\inst{2}
\and\firstname{P.} \lastname{Schellart}\inst{2} 
\and\firstname{O.} \lastname{Scholten}\inst{4\partial } 
\and\firstname{S.} \lastname{Thoudam}\inst{2}
\and\firstname{T.N.G.} \lastname{Trinh}\inst{4, 6}
\and\firstname{S.} \lastname{ter Veen}\inst{5}  (The LOFAR Cosmic Ray KSP)
}

\institute{
	Vrije Universiteit Brussel (Belgium)
\and
	Radboud University Nijmegen (The Netherlands)
\and
	University of California Irvine (USA)
\and
	KVI-CART (The Netherlands)
\and
	ASTRON (The Netherlands)
\and
	University of Groningen (The Netherlands)
	\and
	NIKHEF (The Netherlands)
}

\abstract{%
The low flux of the ultra-high energy cosmic rays (UHECR) at the highest
energies provides a challenge to answer the long standing question about their
origin and nature. A significant increase in the number of detected UHECR is
expected to be achieved by employing Earth's moon as detector, and search for
short radio pulses that are emitted when a particle interacts in the lunar
rock. Observation of these short pulses with current and future radio
telescopes also allows to search for the even lower fluxes of neutrinos with
energies above $10^{22}$ eV, that are predicted in certain Grand-Unifying-Theories
(GUTs), and e.g. models for super-heavy dark matter (SHDM). In this
contribution we present the initial design for such a search with the LOFAR
radio telescope.
}

\maketitle
\section{Introduction}
\label{intro}
The extremely low flux of ultra-high energy cosmic rays (UHECR) makes the
search for their sources a challenging task~(e.g.~\cite{Kotera2011}). The same
challenge arises in testing specific theories of grand-unification~\cite{Bhattacharjee1997} and
super heavy dark matter~\cite{Aloisio2006, Aloisio2015} by their predicted flux of
neutrinos with extreme energies beyond \SI{1}{Z\electronvolt}. To test
these theories and increase the available UHECR data at the highest energies,
detectors several orders of magnitude larger than currently available are
required.
One approach to achieve this necessary increase of detector
area is to use Earth's Moon as detector. The Moon is `read-out' 
by existing radio telescopes that record the
nanosecond radio pulses that are created when a
particle interacts in the lunar rock. 

Several searches with radio telescopes operating in the GHz frequency regime
have so far yielded only upper limits on the  neutrino flux and have not been
sensitive enough to observe the low UHECR flux~\cite{Buitink2010, Hankins1996, James2011, James2007, Bray2014}. The sensitivity of
these searches was in particular limited by the frequency range of the
used telescopes that operate in the GHz regime. While at  GHz frequencies
the expected pulse amplitudes reach a maximum, the pulses cannot escape the moon for most
part of its surface; only pulses from particles hitting the limb are detectable
at Earth. Conversely, similar searches with the LOw Frequency ARray
(LOFAR)~\cite{vanHaarlem2013} can use the full visible lunar surface as
sensitive area, as it operates in a frequency range that allows pulses to
escape the lunar rock in more geometries and is optimal for the lunar detection
of particles~\cite{Scholten2006}.

\section{Signal Processing at LOFAR}
LOFAR consists of several thousand omni-directional antennas located in more than 48
stations distributed throughout European.  Twenty-four of the stations
 are grouped into a dense core of approximately
\SI{2}{\kilo\meter} diameter in the Netherlands.  All stations are equipped with 96 low-band
antennas (LBAs) with a frequency range from~\SIrange{10}{90}{\mega\hertz} and
at least 48 `tiles' of 16 high-band antennas (HBAs) each, with a frequency
range from~\SIrange{110}{240}{\mega\hertz}.  By beamforming, i.e.\ coherent stacking the
signals of the individual antennas with a corresponding delay, the sensitivity
of a station is tuned towards a specific direction. A station thus assumes the
role of a single dish in a classical radio telescope. 

\begin{figure}[tb]
	\centering
	\includegraphics[width=.6\textwidth]{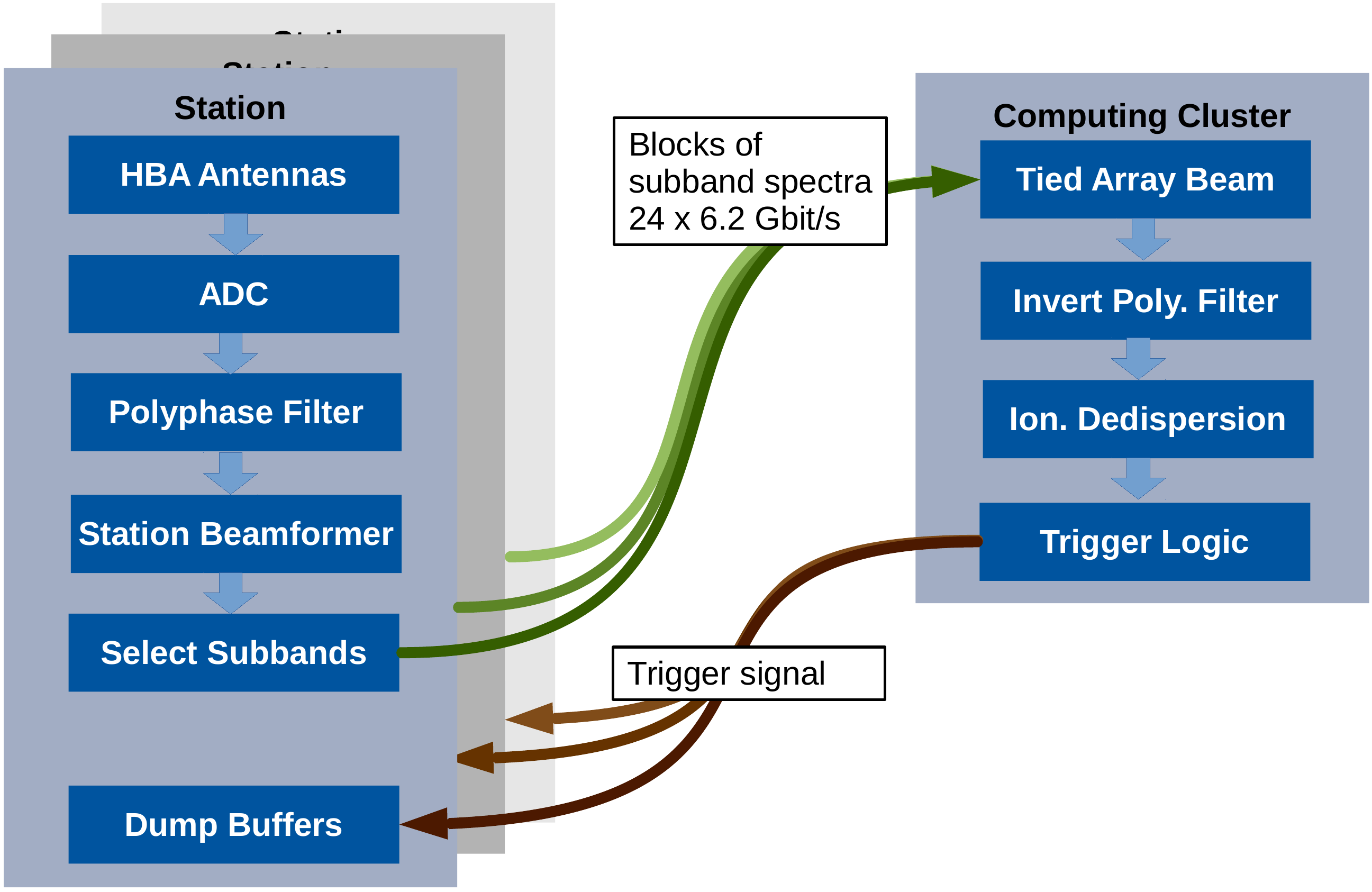}
	\caption{Overview of the online data analysis processing steps for the detection of ns-pulses with LOFAR. }
	\label{fig:ProcessingSteps}
\end{figure}
As preparation for astronomical observations, the data is preprocessed in
several steps in LOFAR before it can be accessed by the user. An overview of
the default processing steps together with the  online-analysis steps required here is given in figure~\ref{fig:ProcessingSteps}.  After
digitization, the data is first processed by a polyphase filter
(PPF), which splits the data stream into 256
sub-channels of \SI{195}{\kilo\hertz} width. Compared to splitting data with a
Fourier transformation, the PPF allows efficient sampling conversion while
preventing signal-leakage between neighboring channels, thus allowing precise
filtering.  However, within the individual sub-channels the time resolution is
reduced from \SI{5}{\nano\second} to only a few micro-seconds due to critical down-sampling.  The filtered
data of the individual antennas of one station is then combined into a `station
beam' before it is transmitted  to a computing cluster  for further processing.

While the pre-filtering of the signal is convenient for typical astronomical
observations, here we require the full nano-second resolution. The thus necessary
inversion of the PPF is a lossy process, but we can use the thereby approximately
reconstructed signal to trigger storage of the raw voltage traces which are
buffered for 5 seconds. We therefore will first combine the
station beams to several tight-array beams pointing to different spots on the
moon, and then construct nanosecond time traces from the combined signals.
These are then corrected for ionospheric dispersion, and eventually trigger a
read-out of the buffer board.

\section{Polyphase Synthesis}
\begin{figure}[b]
	\centering
	\includegraphics[width=\textwidth]{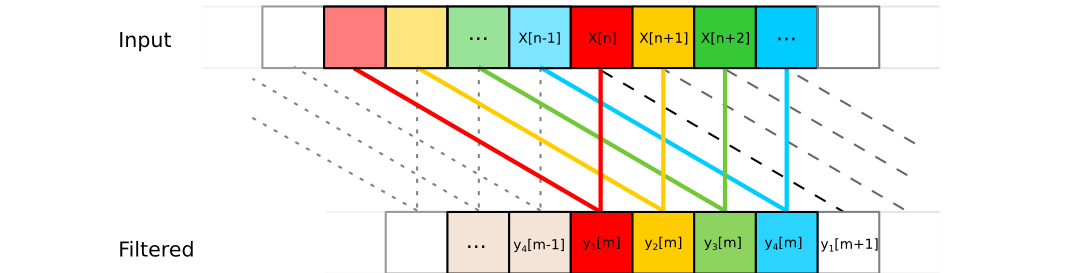}
	\caption{Sketch of the filter stage of an 2-tap 4-channel PPF before application of a 4-point FFT to $\mathbf{y}$.}
	\label{fig:PPFSketch}
\end{figure}

The polyphase filter analysis is an efficient implementation of an multi
channel Finite-Impulse-Response (FIR) filter with corresponding optimization of
the sampling rate to split the signal $x[n]$ into well defined sub-bands
$\hat{y}_i[m]$, with $i = 0 \cdots M$.  The polyphase-filter mathematically
corresponds to the Fourier transform $\mathcal{F}(\mathbf{y})$ of the
transformed signal $\mathbf{y} = H\cdot \mathbf{x}$, with matrix $H$
corresponding to the single-channel filter. For a K-tap filter into
M-sub-channels $H$ is a  $K\cdot M \times M$ matrix. The filter principle is
sketched in figure~\ref{fig:PPFSketch}. 

Inversion of the Fourier transformation in the polyphase analysis is possible
without loss of information. The inverse of the filter matrix $H^{-1}$,
however, does not exists as $H$ is not square.  While methods
exists~(e.g.~\cite{Unser2000}) to construct an approximate inverse $G$ so that
$G\cdot H = 1$, the one-step inversion produces echos of short pulses as e.g.\ in~\cite{Singh2012}. This
limits their usability here, as we thus expect a large number of false triggers
generated from RFI by this method.  We here therefore investigate a more robust
but computationally more complex iterative method to solve the
linear-systems of equations $H\cdot \mathbf{x} = \mathbf{y}$.

As the system is under-determined, a side condition is required to constrain the
solutions. The reason for the ill-determination of the inverse is here, that
any $x[n]$ influences  $y_i[m]$ for $K$ values of $m$. Thus, for a filtered
signal of finite length, $K \times M$ samples at the beginning, respectively
end, of the reconstructed input signal are not completely determined. Searching for 
the minimum solution $\Vert \min{x} \Vert$ can thus be used as side condition
here, as this minimizes the signal in the under-determined areas at cost of  a
slight damping of the reconstructed signal.

An algorithm using this side condition to solve under-determined linear systems
is the LSMR algorithm~\cite{Fong2011}. Compared to alternatives as e.g.~the LSQR algorithm
, the LSMR algorithm has also the advantage of practically monotonically
conversion that allows for safe early termination, which might be advantageous
for a real-time application. 

\begin{figure}[tb]
	\centering
	\includegraphics[width=.82\textwidth]{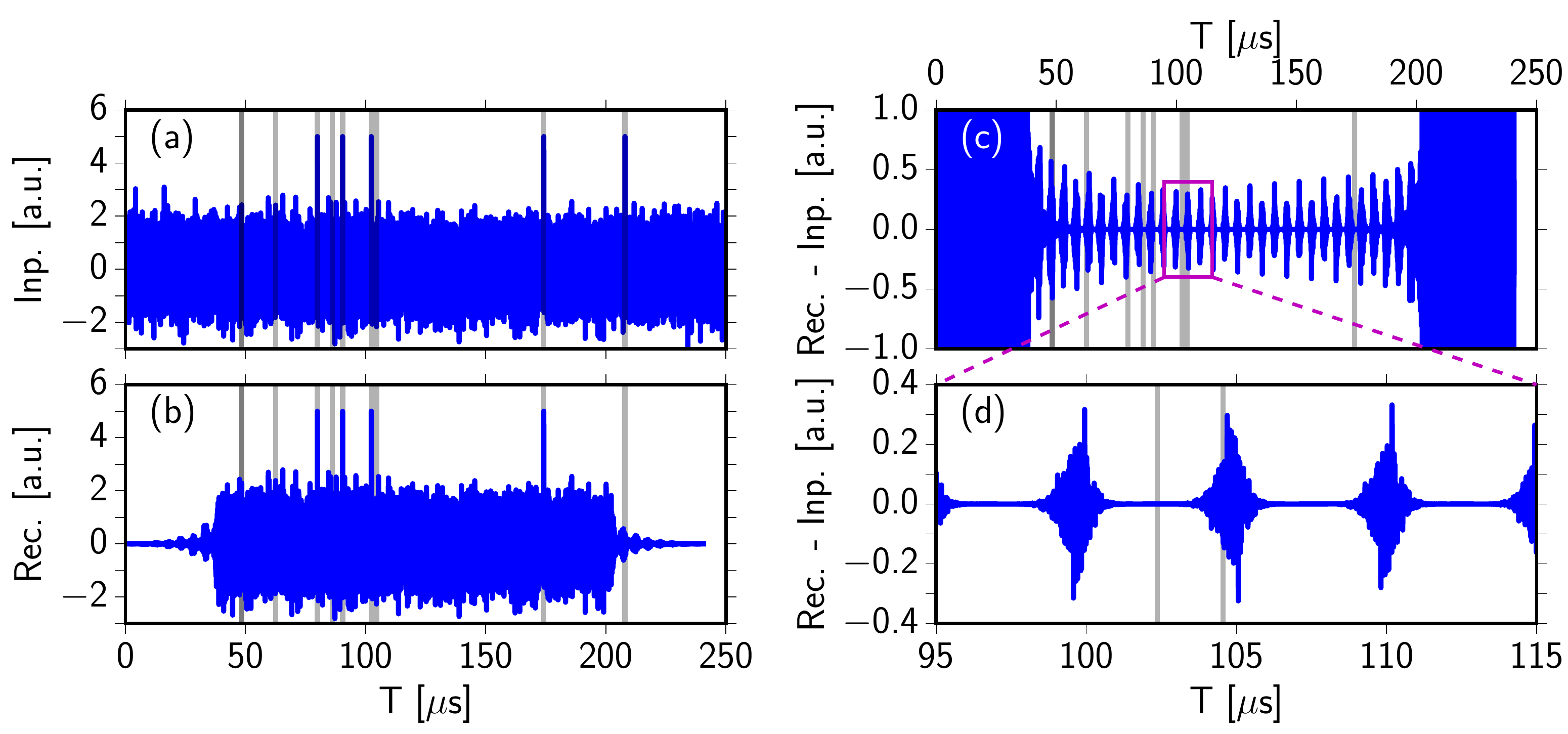}
	\caption{Accuracy of PPF-Synthesis. \textbf{(a)} Input Signal, \textbf{(b)} reconstructed signal, \textbf{(c)}
	difference between input and reconstructed signal, \textbf{(d)} zoom into (c). The position of input pulses are marked by vertical gray lines.}
	\label{fig:PPFExample}
\end{figure}
In figure~\ref{fig:PPFExample} an example of the reconstruction of a trace with
a prototype implementation of this method is given. The input signal shown in
fig.~\ref{fig:PPFExample}~(a) is composed of white noise and delta peaks of
fixed height at random time sets. After applying a polyphase filter and its
inverse to the trace, the signal is recovered almost perfectly, except for the
edges of the trace as shown in fig.~\ref{fig:PPFExample}~(b). The differences
between the input signal and recovered signal  is exactly zero except for small
time windows at intervals corresponding to the number of sub-channels as shown
in fig.~\ref{fig:PPFExample}(c, d). As these windows are known, a necessary increase of the trigger sensitivity is limited to a small time intervall.
  In particular, the uncertainty in the
reconstruction is  not related to the position of the input pulses and thus not
critical for the trigger. The maximum amplitude found here corresponds to less
than 30\% of the root-mean squared of the white noise.

\section{Conclusion}
The LOFAR radio telescope is a promising instrument for the search of
cosmic-particle induced radio-pulses from Earth's moon.  Such a search requires
triggering on preprocessed data and in particular the reconstruction of nano
second resolution time traces via a polyphase synthesis. The iterative approach
to polyphase synthesis investigated here is suitable for triggering the data,
as it does not introduce artefacts that are likely to increase the trigger
rate. 


%
 \bibliography{references}
%
%

\end{document}